\documentclass[preprint2]{aastex62}
\usepackage{subfigure}
\usepackage{url}
\usepackage{comment}
\usepackage{color}
\usepackage{enumitem}
\usepackage{amsmath}

\setlist[enumerate]{noitemsep}
\setlist[enumerate,1]{leftmargin=*}
\setlist[itemize]{noitemsep}
\setlist[itemize,1]{leftmargin=*}
\setlist[description]{noitemsep}
\setlist[description,1]{leftmargin=*}
\setenumerate[0]{label=(\arabic*)}
\shorttitle{RPE stars from CDTGs in the MW Halo}\shortauthors{Agnos et al.}

\newcommand{\cms}[1]{{#1}}

\begin{document}
\title{The $R$-Process Alliance: The $R$-Process Enhancement of Stars from Chemodynamically Tagged Groups in the Milky Way Halo\footnote{This paper includes data gathered with the 6.5~meter Magellan Telescopes located at Las Campanas Observatory, Chile}}
\author{Jessica Merritt Agnos}\affiliation{Department of Physics \& Astronomy, San Francisco State University, San Francisco CA 94132, USA}%\email{jessica.agnos@gmail.com}
\author[0000-0002-5095-4000]{Charli M. Sakari}\affiliation{Department of Physics \& Astronomy, San Francisco State University, San Francisco CA 94132, USA}%\email{sakaricm@sfsu.edu}
\author{Pedro Silva}\affiliation{Department of Physics \& Astronomy, San Francisco State University, San Francisco CA 94132, USA}%\email{psilva3@mail.sfsu.edu}

\author[0000-0001-6154-8983]{Terese T.\ Hansen}%\email{thidemannhansen@gmail.com}
\affiliation{Department of Astronomy, Stockholm University, AlbaNova University Center, SE-106 91 Stockholm, Sweden}

\author[0000-0002-5463-6800]{Erika M.\ Holmbeck}%\email{holmbeck1@llnl.gov}
\affiliation{Lawrence Livermore National Laboratory, 7000 East Avenue, Livermore, CA 94550, USA}
\affiliation{Joint Institute for Nuclear Astrophysics, Center for the Evolution of the Elements (JINA-CEE), East Lansing, MI 48824, USA}

\author[0000-0001-5107-8930]{Ian U.\ Roederer}%\email{iuroederer@ncsu.edu}
\affiliation{Department of Physics and Astronomy, North Carolina State University,
2401 Stinson Dr, Box 8202, Raleigh, NC 27695, USA}
\affiliation{Joint Institute for Nuclear Astrophysics, Center for the Evolution of the Elements (JINA-CEE), East Lansing, MI 48824, USA}

\author{Hal France}\affiliation{Department of Physics \& Astronomy, San Francisco State University, San Francisco CA 94132, USA}%\email{h.france246@gmail.com}

\author{Truman Farr}\affiliation{Department of Physics \& Astronomy, San Francisco State University, San Francisco CA 94132, USA}%\email{trumanfarr@gmail.com}

\author[0000-0002-8504-8470]{Rana Ezzeddine}%\email{rezzeddine@ufl.edu}
\affiliation{Department of Astronomy, University of Florida, Bryant Space Science Center, Gainesville, FL 32611, USA}
\affiliation{Joint Institute for Nuclear Astrophysics, Center for the Evolution of the Elements (JINA-CEE), East Lansing, MI 48824, USA}

\author[0000-0002-2139-7145]{Anna Frebel}%\email{afrebel@mit.edu}
\affiliation{Department of Physics and Kavli Institute for Astrophysics and Space Research, Massachusetts Institute of Technology, Cambridge, MA 02139, USA}
\affiliation{Joint Institute for Nuclear Astrophysics, Center for the Evolution of the Elements (JINA-CEE), East Lansing, MI 48824, USA}

\author[0000-0003-4479-1265]{Vinicius M.\ Placco}%\email{vinicius.placco@noirlab.edu}
\affiliation{NSF NOIRLab, Tucson, AZ 85719, USA}

\author[0000-0003-4573-6233]{Timothy C.\ Beers}%\email{tbeers@nd.edu}
\affiliation{Department of Physics and Astronomy, University of Notre Dame,
225 Nieuwland Science Hall, Notre Dame, IN 46556, USA}
\affiliation{Joint Institute for Nuclear Astrophysics, Center for the Evolution of the Elements (JINA-CEE), East Lansing, MI 48824, USA}

\correspondingauthor{Charli M. Sakari}\email{sakaricm@sfsu.edu}
\begin{abstract}
As part of the ongoing work of the $R$-Process Alliance (RPA), detailed abundance measurements of 29 heavy elements in three metal-poor stars, 2MASS J14592981$-$3852558, 2MASS J19445483$-$4039459, and 2MASS J15211026$-$0607566, are presented based on an analysis of high-resolution ($R\sim 80,000$), high signal-to-noise ``portrait'' spectra from the Magellan Inamori Kyocera Echelle (MIKE) spectrograph on the Magellan-Clay Telescope at Las Campanas Observatory. The selected targets were identified as $r$-process-enhanced metal-poor stars in previous RPA snapshot analyses.  They have also been linked to possible 
chemodynamically tagged groups, indicating that the stars may have formed in dwarf galaxies that were later accreted into the Milky Way halo.   These stars have also been tentatively linked to the Thamnos structure. The detailed chemical abundances in this work confirm that 2MASS J14592981$-$3852558 and J15211026$-$0607566 are $r$-II stars, while 2MASS J19445483$-$4039459 is found to lie just below the threshold for $r$-I status. The $r$-II stars show signs of slight enhancement in fission fragments compared to 2MASS J19445483$-$4039459.  Based on radioactive age dating with Th, the $r$-process material in the two $r$-II stars is found to be old (with ages $>10$ Gyr); neither star shows signs of an actinide boost. The varying elemental compositions suggest that these stars likely did not originate in the same environment, though each could be consistent with originating in the Thamnos progenitor.
\end{abstract}  

\section{Introduction}\label{sec:Intro}
There is still uncertainty surrounding the nucleosynthesis of many of the heavier elements (Z$>$30), including the astrophysical sites and the physics of the high energy events that form them. The heaviest elements are created through the rapid ($r$-) neutron-capture process \citep{1957RvMP...29..547B}. Metal-poor stars ([Fe/H]$\lesssim -$1; \citealt{2005Beers,2018ARNPS..68..237F}) that are enriched in $r$-process elements such as Eu provide valuable constraints on early $r$-process nucleosynthesis and the early assembly history of the Milky Way (MW). Even now, while we have seen direct evidence of $r$-process nucleosynthesis in a neutron star merger (NSM; \citealt{2022Curtis}), there is uncertainty as to how these events, in the known timescale of the universe, were able to produce the amount of $r$-process elements we observe. Measurement of the chemical abundances of $r$-process-enhanced metal-poor stars is important to constrain early $r$-process nucleosynthesis and help to understand their formation sites in the MW. 

In its first five data releases (\citealt{2018ApJ...868..110S,2018ApJ...858...92H,2020ApJS..249...30H,2020ApJ...898..150E,2024ApJS..274...39B}), the $R$-Process Alliance (RPA) has discovered hundreds of $r$-process enhanced metal-poor stars in the MW.  The RPA was formed in 2017 with the goal of increasing the number of known $r$-process enhanced metal-poor stars to study the sites of $r$-process nucleosynthesis, investigate the chemical evolution of the MW, and motivate the need for additional nuclear data.\footnote{https://sites.google.com/view/rprocessalliance/home} To identify new $r$-process enhanced metal-poor stars, the RPA has a multi-step observing strategy.  Phase I (e.g., \citealt{2018AJ....155..256P,2019ApJ...870..122P}) involves identifying bright, very metal-poor targets from large surveys and confirming they are metal-poor through medium-resolution spectroscopy.  During Phase II (\citealt{2018ApJ...868..110S,2018ApJ...858...92H,2020ApJS..249...30H,2020ApJ...898..150E,2024ApJS..274...39B}), ``snapshot'' high-resolution spectra ($R\sim$ 25,000$-$35,000; S/N $\sim$ 30) are obtained for confirmed metal-poor stars to determine Sr, Ba, and Eu abundances and identify stars with $r$-process enrichment.  Phase III (e.g., \citealt{2018ApJ...854L..20S,2018ApJ...859L..24H,2024A&A...688A.123X,2025A&A...704A.282R}) involves obtaining higher-resolution ``portrait'' spectra ($R\gtrsim 50,000$; S/N $\geq$ 50) of particularly interesting $r$-process enhanced stars.  \cms{These higher-quality portrait spectra go further into the blue spectral region, at higher resolution, so that more absorption features can be measured and abundances for more elements can be determined.}

This work presents detailed analyses of portrait spectra for three stars:\\ 2MASS J14592981$-$3852558 (J1459$-$3852), 2MASS J15211026$-$0607566 (J1521$-$0607), and 2MASS J19445483$-$4039459 (J1944$-$4039).  These stars were confirmed to be metal-poor by \cite{2018AJ....155..256P}. \cite{2018ApJ...868..110S} followed up with Phase II snapshot spectra for J1521$-$0607, and \cite{2020ApJS..249...30H} did the same for J1459$-$3852 and J1944$-$4039. The three targets were all found to be $r$-process enhanced, which warranted additional portrait spectra \cms{at higher resolution and S/N ratio}.  These three stars have also been kinematically linked to other stars, which means that they are each part of a retrograde chemodynamically tagged group (CDTG) or a dynamically tagged group (DTG).  The stars may or may not be associated with the same CDTG/DTG, since classifications vary between papers (e.g., \citealt{2021ApJ...908...79G, 2023ApJ...943...23S, 2023ApJ...946...48H}).  These CDTGs may be remnants of accreted satellite galaxies or disrupted star clusters.  Analyses of these portrait spectra \cms{in this paper} will reveal important information about early $r$-process nucleosynthesis in the progenitors of these CDTGs, and may shed light on the early formation of the MW halo.

The remainder of this paper is arranged as follows.  The observations, data reduction, and stellar parameters are discussed in Section \ref{sec:datatargets}.  Section \ref{sec:ChemAbun} describes the spectral synthesis analysis techniques. The final abundances are presented and discussed in Section \ref{sec:results}, and the results are summarized in Section \ref{sec:summary}.

\section{Observations and Targets}\label{sec:datatargets}

\subsection{Observations, Data Reduction, and Stellar Parameters}\label{sub:obsinfo}
The stars were observed in 2017 and 2018, using the Magellan-Clay Telescope at Las Campanas Observatory, with the Magellan Inamori Kyocera Echelle (MIKE) spectrograph \citep{2003SPIE.4841.1694B}. The slit size and binning for each target are shown in Table \ref{table:info}. For blue wavelengths (3340-5050~{\AA}), the predicted resolving power is $R \sim$ 83,000. For the red wavelengths (4850-9400~{\AA}), the predicted resolving power is $R \sim$ 65,000.  Basic information about the program stars, including total exposure times, are shown in Table \ref{table:info}.

The data were reduced using the Carnegie Python (CarPy) pipeline \citep{kelson2000,kelson2003}. The individual spectra were then normalized in IRAF using polynomial fits to each order's blaze function, the spectra were shifted to the rest frame, and individual visits were combined.  Radial velocities were determined through cross-correlations with a template spectrum. Final heliocentric radial velocities are given in Table \ref{table:info}, along with final S/N ratios of the combined spectra.  The measured heliocentric radial velocities are in excellent agreement with those from {\it Gaia} DR3 \citep{2023A&A...674A...1G}.  Additional radial velocity monitoring by the RPA with Magellan and MIKE also found no signs of significant variations.\footnote{J1459$-$3852 was observed again on 2022 March 3 and 2023 January 29, with $v_{\rm{helio}}=302.9$ and $302.0$ km~s$^{-1}$, respectively.  J1521$-$0607 was observed on 2022 March 3 and 2023 January 29, with $v_{\rm{helio}}=-11.8$ and $-13.8$ km~s$^{-1}$, respectively.  J1944$-$4039 was observed again on 2022 March 3, with $v_{\rm{helio}}=99.2$ km~s$^{-1}$.}

Table \ref{table:info} lists the basic information and stellar parameters of each of the three stars being analyzed. The parameters of all three stars were adopted from the upcoming RPA snapshot data release of $\sim 2000$ (2K) stars (Hansen et al., \cms{{\it in prep.}).\footnote{\cms{Note that the stellar parameters for the snapshot spectra \citep{2018ApJ...868..110S,2020ApJS..249...30H} were determined in different ways.  The adoption of the 2K parameters ensures that the stars are being analyzed in a consistent way.}}}  A similar procedure has been used in previous RPA papers (e.g., \citealt{2024ApJ...971..158R,2024MNRAS.529.1917S,2025A&A...704A.282R}). Briefly, effective temperatures are determined from empirical calibrations using {\it Gaia} and 2MASS photometry, surface gravities are determined from fundamental relations using {\it Gaia} DR3 parallaxes \citep{2023A&A...674A...1G}, using the same procedure described in \citet{2018ApJ...865..129R}.  Microturbulent velocities are derived from an empirical relationship with surface gravity, and the metallicity comes from the average abundance of the \ion{Fe}{1} lines.  More details on the stellar parameter determination will be discussed in Hansen et al. ({\it in prep.}). The corrected C abundances of the stars ($[\rm{C/Fe}]<+0.7$; \citealt{2018ARNPS..68..237F}) show that they are not carbon-enhanced metal-poor (CEMP) stars.

\begin{deluxetable*}{@{}lccc}
\tabletypesize{\footnotesize}
\tablecolumns{4}
\tablewidth{0pt}
\tablecaption{Basic Target Information\label{table:info}}
%\hspace*{-4in}
\tablehead{
\colhead{Star ID\tablenotemark{a}:} & \colhead{J14592981$-$3852558} & \colhead{J15211026$-$0607566} & \colhead{J19445483$-$4039459} 
}
\startdata
RA & 14:59:29.81 & 15:21:10.33 & 19:44:54.83 \\
Dec & $-$38:52:55.8 & $-$06:07:56.5 & $-$40:39:45.9 \\
$V$ mag& 11.08 & 11.24 & 10.58 \\
Observing Dates & 28 Feb, 1 March, 2018 & 9 May, 2017 & 25 July, 2018 \\
Observing Setup & $0.35\arcsec$ slit, 2x1 binning & $0.35\arcsec$ slit,  2x2 binning & $0.5\arcsec$ slit, 2x1 binning\\
Integration Time (s) & 12000 & 4800 & 9000 \\
S/N$^{b}$ (4000 \AA) & 50  & 34 & 25 \\
S/N$^{b}$ (6000 \AA) & 403 & 297 & 387 \\
$v_{\rm{helio}}$ (km s$^{-1}$) & $302.3\pm0.5$ (night 1) & $-12.9\pm0.5$ & $98.5\pm0.50$\\
 & $302.1\pm0.5$ (night 2) & & \\
T$_{\rm{eff}}$ (K)$^{c}$  & 4336 & 4894 & 4437 \\
$\log g^{c}$              & 0.57 & 1.59 & 0.96 \\
$\xi$ (km s$^{-1}$)$^{c}$ & 2.49 & 1.88 & 2.24 \\
%[$M/H$]^{c}$           & $-$2.51 & $-$1.98 & $-$1.99 \\
$[$\ion{Fe}{1}/H$]^{c}$         & $-2.56\pm0.17$ & $-1.91\pm0.16$ & $-2.14\pm0.10$ \\
$[$\ion{Fe}{2}/H$]^{c}$         & $-2.52\pm0.12$ & $-2.07\pm0.07$ & $-2.04\pm0.17$ \\
%Uncorrected 2K [C/Fe] & $-0.83\pm-0.05$ & $0.04\pm0.05$ & $-0.36\pm0.06$ \\
$[$C/Fe$]$$^{c,d}$ & $-0.05\pm0.05$ & $0.41\pm0.05$ & $0.30\pm0.06$ \\
Class$^{c}$ & $r$-II & $r$-II  & None \\
CDTG Group: \cite{2021ApJ...908...79G} & CDTG-2 & CDTG-27 & CDTG-2\\
CDTG Group: \citet{2023ApJ...943...23S} & None   & CDTG-22 & None \\
DTG Group: \citet{2023ApJ...946...48H} & DTG-1 & DTG-16 & n/a \\
\enddata
\tablenotetext{a}{The stars are identified by their IDs from the Two Micron All Sky Survey (2MASS; \citealt{2006AJ....131.1163S}).}
\tablenotetext{b}{S/N is per pixel.}
\tablenotetext{c}{The parameters derived in the RPA 2K snapshot sample are adopted (Hansen et al., {\it in prep.}).}
%\from the snapshot spectra of \cite{2018ApJ...868..110S} and \cite{2020ApJS..249...30H} are adopted.}
\tablenotetext{d}{This is the corrected [C/Fe], accounting for C depletion along the red giant branch, following \citet{Placco2014}.}
\end{deluxetable*}

\clearpage
\subsection{Associated Chemodynamically Tagged Groups}\label{sub:assocCDTG}
The stars were selected for Phase III portrait observations because they were identified as $r$-process enhanced stars.  They were subsequently found to \cms{have low-energy, retrograde orbits, indicating an ancient accretion origin from a dwarf galaxy.  Clustering studies found that the stars were clumped in kinematic space, indicating they could be part of CDTGs.} As discussed in Section \ref{sec:Intro}, CDTGs may be remnants of accreted dwarf galaxies.  \cms{Detailed chemical abundance studies can shed light on the properties of the birth site(s) of these stars, including whether they likely formed in the same place.}

\cite{2021ApJ...908...79G} first assigned these targets to CDTGs, with J1459$-$3852 and J1944$-$4039 being members of CDTG-2 and J1521$-$0607 a member of CDTG-27. \citet{2021ApJ...908...79G} compared their CDTGs to known substructures and the DTGs of \citet{2020ApJ...891...39Y}  and \citet{2021ApJ...907...10L}, which were all grouped without any knowledge of the chemistry of their member stars.  \cms{They found that both CDTG-2 and CDTG-27 could be associated with the Thamnos substructure, a known stream that has been associated with an ancient accretion event \citep{2019A&A...631L...9K}.}  \citet{2023ApJ...943...23S} also linked J1521$-$0607 to a CDTG associated with the Thamnos substructure.  If these three stars are indeed associated with the Thamnos substructure, the portrait analyses in this paper could provide valuable insight into its chemical evolution.

\section{Spectrum Synthesis and Chemical Abundances}\label{sec:ChemAbun}
The list of spectral lines to be analyzed in this work was assembled by combining the lines used in \cite{2017ApJ...844...18P}, \cite{2018ApJ...868..110S}, and Hansen et al., (in preparation) as a starting point for 29 neutron-capture elements: \ion{Sr}{2}, \ion{Y}{2}, \ion{Zr}{2}, \ion{Mo}{1}, \ion{Ru}{1}, \ion{Rh}{1}, \ion{Pd}{1}, \ion{Ag}{1}, \ion{Ba}{2}, \ion{La}{2}, \ion{Ce}{2}, \ion{Pr}{2}, \ion{Nd}{2}, \ion{Sm}{2}, \ion{Eu}{2}, \ion{Gd}{2}, \ion{Tb}{2}, \ion{Dy}{2}, \ion{Ho}{2}, \ion{Er}{2}, \ion{Tm}{2}, \ion{Yb}{2}, \ion{Lu}{2}, \ion{Hf}{2}, \ion{Os}{1}, \ion{Ir}{1}, \ion{Pb}{1}, \ion{Th}{2}, and \ion{U}{2}.  The lines spanned wavelengths from approximately 3380~{\AA} to approximately 7950~{\AA}. Lines were excluded from the analysis if they were undetectable above the noise, significantly blended, or saturated.  The analyzed lines and the abundances for each star are given in Table \ref{table:lineabundances}.

\cms{The selected lines add many additional elements that were not measured in the snapshot analyses, particularly the lighter neutron-capture elements Mo, Ru, Rh, Pd, and Ag and the heavier elements Tb, Ho, Er, Tm, Yb, Lu, Hf, Ir, and U.  The snapshot analysis of J1521$-$0607 included abundances of 13 neutron-capture elements \citep{2018ApJ...868..110S}--this portrait analysis has added detections of 14 new elements and one upper limit.  The snapshot analyses of J1459$-$3852 and J1944$-$4039 \citep{2020ApJS..249...30H} only presented measurements of Sr, Ba, and Eu.  For J1459$-$3852, this portrait analysis presents measurements of 23 new elements and two upper limits, while measurements of 21 new elements and two upper limits are presented for J1944$-$4039.}

 %abundances Table
\begin{deluxetable*}{lccccccccc}     
%{ld{3}cd{3}cd{3}c}
%\newcolumntype{d}[1]{D{,}{\;\pm\;}{#1}}
%  \begin{tabular}{@{}ld{6}ccd{6}cd{6}@{}}
\tabletypesize{\footnotesize}
\tablecolumns{10}
\tablewidth{0pt}
\tablecaption{Individual Line Abundance Measurements for Target Stars
\label{table:lineabundances}}
\tablehead{
 \colhead{Element} & $\lambda$ (\AA) & EP (eV) & $\log gf$ & \multicolumn{2}{c}{J1459$-$3852} & \multicolumn{2}{c}{J1521$-$0607} & \multicolumn{2}{c}{J1944$-$4039} \\
 & & & & $\log \epsilon$ & $\sigma$ & $\log \epsilon$ & $\sigma$ & $\log \epsilon$ & $\sigma$
}
\startdata
\ion{Sr}{2} & 4161.792 & 2.940 & $-$0.47 &    0.88 & 0.10 & 1.14 & 0.15 & 0.98 & 0.20 \\
\ion{Y}{2}  & 3747.556 & 0.104 & $-$0.91 & $-$0.05 & 0.10 & 0.21 & 0.25 & $-$0.08 & 0.20 \\
\ion{Y}{2}  & 4398.013 & 0.129 & $-$1.00 & $-$0.20 & 0.10 & 0.39 & 0.20 & 0.02 & 0.20 \\
\ion{Y}{2}  & 4682.324 & 0.408 & $-$1.51 & $-$0.20 & 0.10 & 0.46 & 0.20 & 0.10 & 0.20 \\
\ion{Y}{2}  & 4883.684 & 1.083 &    0.07 & $-$0.37 & 0.10 & 0.28 & 0.20	& \nodata & \nodata \\	
\ion{Y}{2}  & 4900.12  & 1.032 & $-$0.09 & $-$0.33 & 0.10 & 0.28 & 0.20 & $-$0.03 & 0.15 \\
\ion{Zr}{2} & 3836.761 & 0.559 & $-$0.12 &   0.65 & 0.10 & \nodata & \nodata & 0.59 & 0.15 \\
\ion{Zr}{2} & 3991.127 & 0.758 & $-$0.31 &   0.62 & 0.10 & \nodata & \nodata & 0.64 & 0.20 \\
\ion{Zr}{2} & 3998.954 & 0.559 & $-$0.52 &   0.70 & 0.10 & 1.00 & 0.25 & 0.79 & 0.20 \\
\ion{Zr}{2} & 4050.316 & 0.713 & $-$1.06 &   0.59 & 0.10 & 1.15 & 0.20 & 0.84 & 0.20 \\
\enddata
\tablecomments{(This table is available in its entirety in machine-readable form.)}
\end{deluxetable*}

\begin{figure*}[h!]
\centering\hspace*{-0.25in}
    \includegraphics[scale=0.64,trim=0.9in 0.2in 1.2in 0.4in,clip]{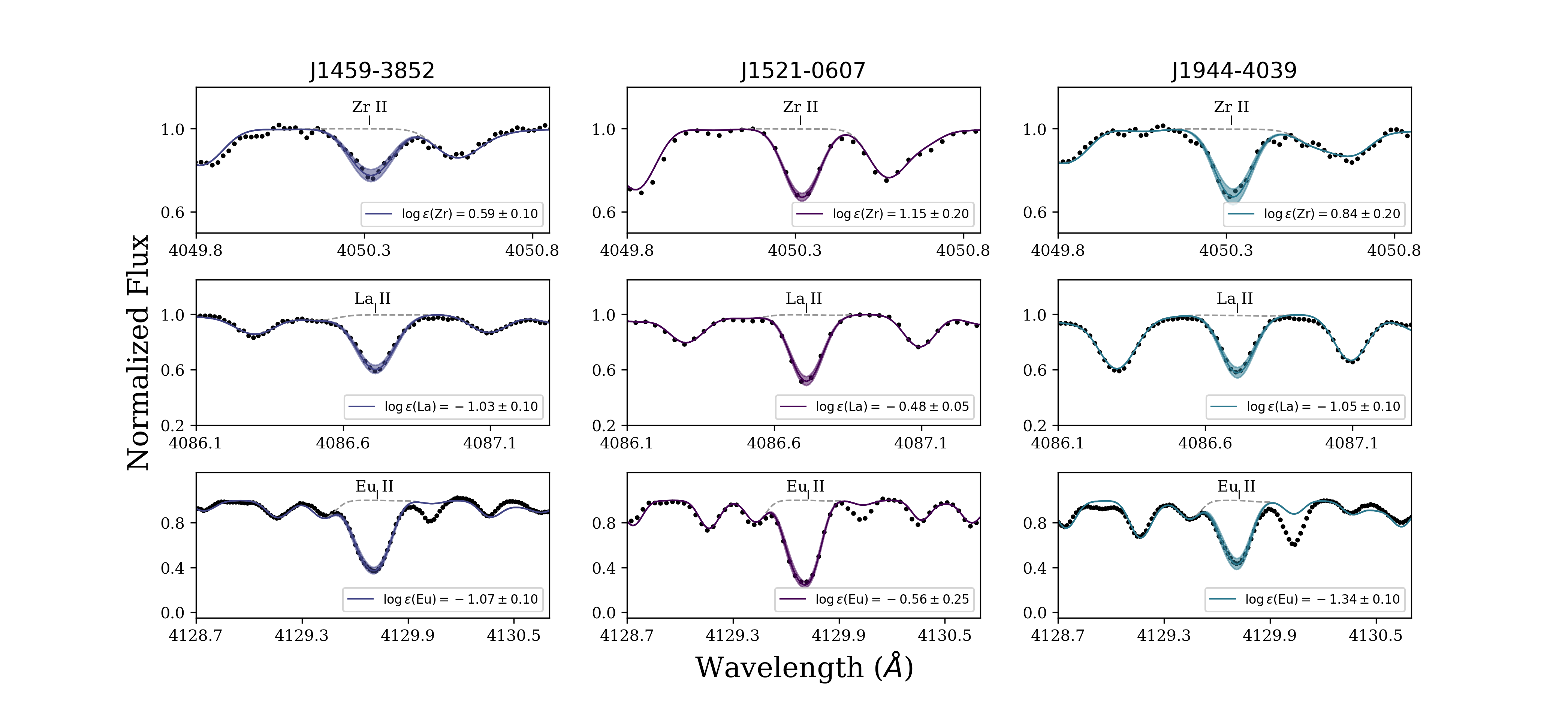}
    \caption{Example syntheses of the 4050.3~{\AA} \ion{Zr}{2} line (top row), the 4086.7~{\AA} \ion{La}{2} line (middle row), and the 4129.7~{\AA} \ion{Eu}{2} line (bottom row) in J1459$-$3852 (left column), J1521$-$0607 (middle column), and J1944$-$4039 (right column), respectively.  The black points show the spectrum, the solid lines show the best-fit syntheses, and the shaded regions show abundance uncertainties.  The dashed gray lines show the synthesis with zero abundance for the line of interest.}
    \label{fig:Synths}
\end{figure*}

The {\tt linemake} code\footnote{\url{https://github.com/vmplacco/linemake}}  \citep{placco2021,2021ascl.soft04027P} was used to create lists of all lines around the line of interest. These lists include hyperfine structure, isotopic splitting, and molecular features due to CH, C$_{2}$, and CN. The $r$-process isotopic ratios of \citet{2008ARA&A..46..241S} were adopted.  Lines of interest were synthesized with the 2017 version of the local thermodynamic equilibrium (LTE) line-analysis code, \texttt{MOOG} \citep{1973ApJ...184..839S}.  This version of {\tt MOOG} treats Rayleigh scattering as isotropic, coherent scattering, as described in \citet{2011AJ....141..175S}.\footnote{\url{https://github.com/alexji/moog17scat}}  %The stellar parameters (effective temperature, T$_{\rm{eff}}$; surface gravity, $\log g$; microturbulent velocity, $\xi$; and metallicity, [Fe/H]) and [C/Fe] ratios determined 
The best-fitting abundances were found by minimizing the residuals between the observed and synthesized spectra.  Uncertainties in the abundance of each spectral line were determined on the basis of the residuals, considering the noise level and nearby blends. The upper limits for some elements were determined by finding the maximum allowed abundance for an undetected spectral line given the noise level.  Example syntheses of Zr, La, and Eu lines are shown in Figure \ref{fig:Synths}.

The final $\log \epsilon$ abundances\footnote{The standard notation is used:\\ $\log \epsilon = \log(N_{\rm{X}}/N_{\rm{H}}) + 12$, where $N_{\rm{X}}$ is the number density of any element X.} and uncertainties for each line are listed in Table \ref{table:lineabundances}, along with the atomic data for each transition.

The mean $\log \epsilon$ and [X/Fe] abundances for each element are shown in Table \ref{table:allabundances}. Uncertainties for the mean log~$\epsilon$ abundances were calculated in two ways: for elements with multiple lines, the line-to-line standard deviation was adopted, with a minimum uncertainty of 0.05 dex, while for elements with a single line, the line uncertainty was adopted.

\cms{Abundance uncertainties due to the uncertainties in the stellar parameters are shown in Table \ref{table:sysuncertainties}.  Representative uncertainties of $\pm 150$\;K were adopted for the effective temperature, $0.2$ for $\log g$, $0.2$ km s$^{-1}$ for the microturbulent velocity, and 0.1 dex for [M/H].}

 %abundances Table
\begin{deluxetable*}{lrccccrccccrccc}     
%{ld{3}cd{3}cd{3}c}
%\newcolumntype{d}[1]{D{,}{\;\pm\;}{#1}}
%  \begin{tabular}{@{}ld{6}ccd{6}cd{6}@{}}
\tabletypesize{\footnotesize}
\tablecolumns{12}
\tablewidth{0pt}
\tablecaption{Final Mean Neutron-Capture Abundance Measurements for Target Stars
\label{table:allabundances}}
\tablehead{
 \colhead{Element} & \multicolumn{4}{c}{J1459$-$3852} & & \multicolumn{4}{c}{J1521$-$0607} & & \multicolumn{4}{c}{J1944$-$4039} \\
       \colhead{}  & \colhead{$\log \epsilon$(X)} & \colhead{$\sigma$}& \colhead{N\tablenotemark{a}} & \colhead{[X/Fe]\tablenotemark{b}} & & \colhead{$\log\epsilon$(X)} & \colhead{$\sigma$}& \colhead{N\tablenotemark{a}} & \colhead{[X/Fe]\tablenotemark{b}} & & \colhead{$\log \epsilon$(X)} & \colhead{$\sigma$}& \colhead{N\tablenotemark{a}} & \colhead{[X/Fe]\tablenotemark{b}} 
}
\startdata
Sr II & $+0.88$ & $0.10$ & 1  & $+0.53$ & & $+1.14$ & $0.15$  & 1  & $+0.34$ & & $+0.98$ & $0.20$  & 1 & $+0.15$  \\
Y II  & $-0.23$ & $0.13$ & 5  & $+0.08$ & & $+0.32$ & $0.09$  & 6  & $+0.18$ & & $0.00$ & $0.09$  & 4 & $-0.17$  \\
Zr II & $+0.61$ & $0.06$ & 7  & $+0.55$ & & $+1.13$ & $0.07$  & 7  & $+0.62$ & & $+0.72$ & $0.12$  & 4 & $+0.18$  \\
Mo I  & $-0.08$ & $0.20$ & 1  & $+0.60$ & & $+0.58$ & $0.20$  & 1  & $+0.61$ & & $+0.12$ & $0.15$  & 1 & $+0.38$  \\
Ru I  & $+0.12$ & $0.14$ & 2  & $+0.93$ & & $+0.74$ & $0.16$  & 3  & $+0.90$ & & $+0.06$ & $0.14$  & 2 & $+0.45$  \\
Rh I  & $-0.80$ & $0.20$ & 1  & $+0.85$ & & $-0.02$ & $0.25$ & 1  & $+0.98$ & & $-0.88$ & $0.20$ & 1 & $+0.35$  \\
Pd I  & $-0.24$ & $0.09$ & 2  & $+0.75$ & & $+0.21$ & $0.06$ & 3  & $+0.55$ & & $-0.42$ & $0.08$ & 3 & $+0.150$  \\
Ag I  & $-1.07$ & $0.30$ & 1  & $+0.55$ & & $-0.79$ & $0.30$ & 1  & $+0.18$ & &  \nodata &       & 0 & \nodata \\
Ba II & $-0.21$ & $0.07$ & 3  & $+0.13$ & & $+0.48$ & $0.16$  & 2 & $+0.37$ & & $-0.05$ & $0.06$ & 3  & $-0.19$  \\
La II & $-1.04$ & $0.10$ & 10 & $+0.38$ & & $-0.39$ & $0.07$ & 7  & $+0.58$ & & $-0.97$ & $0.15$ & 4  & $-0.03$  \\
Ce II & $-0.68$ & $0.07$ & 13 & $+0.26$ & &  $0.00$ & $0.06$ & 15 & $+0.49$ & & $-0.60$ & $0.07$ & 10 & $-0.14$ \\
Pr II & $-1.17$ & $0.14$ & 5  & $+0.63$ & & $-0.49$ & $0.11$ & 6  & $+0.86$ & & $-1.17$ & $0.29$ & 3  & $+0.15$  \\
Nd II & $-0.57$ & $0.10$ & 19 & $+0.53$ & & $+0.11$ & $0.09$ & 18 & $+0.76$ & & $-0.58$ & $0.08$ & 12 & $+0.04$ \\
Sm II & $-0.86$ & $0.06$ & 3  & $+0.70$ & & $-0.18$ & $0.09$ & 7 & $+0.93$ & & $-0.88$ & $0.11$ & 3  & $+0.20$  \\
Eu II & $-1.15$ & $0.13$ & 5  & $+0.85$ & & $-0.56$ & $0.09$ & 6 & $+0.99$ & & $-1.29$ & $0.06$ & 6  & $+0.23$ \\
Gd II & $-0.57$ & $0.10$ & 8  & $+0.88$ & & $-0.10$ & $0.18$ & 8 & $+0.90$ & & $-0.87$ & $0.08$ & 6  & $+0.10$  \\
Tb II & $-1.34$ & $0.09$ & 2  & $+0.88$ & & $-0.86$ & $0.08$ & 4 & $+0.91$ & & $-1.59$ & $0.20$ & 1  & $+0.15$  \\
Dy II & $-0.44$ & $0.12$ & 7  & $+0.98$ & & $+0.13$ & $0.12$ & 6 & $+1.10$ & & $-0.66$ & $0.11$ & 3  & $+0.28$  \\
Ho II & $-1.03$ & $0.07$ & 2  & $+1.01$ & & $-0.59$ & $0.15$ & 4 & $+1.00$ & & $-1.41$ & $0.30$ & 1  & $+0.15$  \\
Er II & $-0.63$ & $0.12$ & 5  & $+0.97$ & & $-0.11$ & $0.14$ & 6 & $+1.04$ & & $-0.99$ & $0.13$ & 5  & $+0.13$  \\
Tm II & $-1.46$ & $0.07$ & 5  & $+0.96$ & & $-0.95$ & $0.13$ & 5 & $+1.02$ & & $-1.73$ & $0.12$ & 3  & $+0.21$  \\
Yb II & $-0.72$ & $0.15$ & 1  & $+0.96$ & & $-0.54$ & $0.40$ & 1 & $+0.69$ & & $-1.10$ & $0.20$ & 1  & $+0.10$  \\
Lu II & \nodata &\nodata & 0 & \nodata & & $-0.78$ & $0.15$ & 1  & $+1.19$ & & $-1.39$ & $0.40$ & 1  & $+0.55$  \\
Hf II & $-0.96$ & $0.35$ & 3  & $+0.71$ & & $-0.36$ & $0.13$ & 2 & $+0.86$ & & $-1.14$ & $0.40$ & 1  & $+0.05$  \\
Os I  & $-0.31$ & $0.10$ & 1  & $+0.85$ & & $+0.52$ & $0.11$ & 2 & $+1.03$ & & $-0.39$ & $0.20$ & 1 & $+0.35$  \\
Ir I  & $-0.05$ & $0.20$ & 1  & $1.13$ & & $+0.48$ & $0.30$  & 1 & $+1.07$ & &  \nodata &       & 0 & \nodata  \\
Pb I  & $<-0.03$ &       & 1 & $<+0.78$ & & $<-0.03$ &       & 1 & $<+0.13$ & & $<+0.08$ & & 1  & $<+0.47$ \\
Th II & $-1.69$ & $0.15$ & 1  & $+0.81$ & & $-1.04$ & $+0.20$ & 1  & $+1.01$ & & $<-1.50$ & & 1 & $<+0.52$  \\
U II  & $<-2.20$ &       & 1 & $<+0.86$ & &  \nodata &      & 0  & \nodata & &  \nodata  &      & 0 & \nodata  \\
& & & & & & & & & & & & & \\
$[$Ba/Eu$]$ &  & & & $-$0.72 & & & & & $-$0.62 & & & & &  $-$0.42 \\
$[$Sr/Ba$]$ &  & & & $+$0.40 & & & & & $-$0.03 & & & & &  $+$0.33 \\
\enddata
\tablenotetext{a}{Number of spectral lines used to calculate mean abundance}
\tablenotetext{b}{For neutral species, [X/Fe] is calculated with [\ion{Fe}{1}/H], while for singly ionized species [\ion{Fe}{2}/H] is used.}
\end{deluxetable*}

 %abundances Table
\begin{deluxetable*}{llcccccccccccc} 
%{ld{3}cd{3}cd{3}c}
%\newcolumntype{d}[1]{D{,}{\;\pm\;}{#1}}
%  \begin{tabular}{@{}ld{6}ccd{6}cd{6}@{}}
\tabletypesize{\scriptsize}
\tablecolumns{18}
\tablewidth{0pt}
\tablecaption{Uncertainties in Mean Abundance Measurements from Atmospheric Parameters
\label{table:sysuncertainties}}
\tablehead{
& & \multicolumn{4}{c}{J1459$-$3852}  & \multicolumn{4}{c}{J1521$-$0607}  & \multicolumn{4}{c}{J1944$-$4039}\\
& & \multicolumn{4}{c}{$\Delta \log \epsilon$(X)}  & \multicolumn{4}{c}{$\Delta \log \epsilon$(X)}  & \multicolumn{4}{c}{$\Delta \log \epsilon$(X)}\\
&  & \multicolumn{1}{c}{$T_{\rm{eff}}$ (K)} & \multicolumn{1}{c}{$\log g$} & \multicolumn{1}{c}{$\xi$ (km s$^{-1}$)} & \multicolumn{1}{c}{[M/H]} & \multicolumn{1}{c}{$T_{\rm{eff}}$ (K)} & \multicolumn{1}{c}{$\log g$} & \multicolumn{1}{c}{$\xi$ (km s$^{-1}$)} & \multicolumn{1}{c}{[M/H]} & \multicolumn{1}{c}{$T_{\rm{eff}}$ (K)} & \multicolumn{1}{c}{$\log g$} & \multicolumn{1}{c}{$\xi$ (km s$^{-1}$)} & \multicolumn{1}{c}{[M/H]}\\
& & $\pm150$ & $\pm0.2$ & $\pm0.2$ & $\pm0.1$ & $\pm150$ & $\pm0.2$ & $\pm0.2$ & $\pm0.1$ & $\pm150$ & $\pm0.2$ & $\pm0.2$ & $\pm0.1$ }
\startdata
\ion{Sr}{2} & $+$\tablenotemark{a} & ${\bf -0.1}$ & $+0.02$ & $-0.01$ & $0$ & 
$+0.06$  & $+0.05$ & $-0.05$ & $+0.03$ & 
$-0.05$  & $+0.05$ & $-0.01$ & $+0.01$ \\
           & $-$ & ${\bf +0.13}$  & $0.0$ & $+0.01$ & $+0.01$ & 
           $-0.05$  & $-0.03$ & $+0.04$ & $-0.01$ & 
           $+0.06$  & $-0.03$ & $+0.01$ & $0.0$ \\
\ion{Y}{2} & $+$ & $0$ & $+0.02$ & $-0.05$ & $+0.01$ & 
$+0.07$  & $+0.07$ & $-0.07$ & $+0.03$ & 
$+0.03$  & $+0.05$ & $-0.05$ & $+0.01$ \\
           & $-$ & $+0.01$ & $0.0$ & $+0.05$ & $0$ & 
$-0.07$  & $-0.06$ & $+0.08$ & $-0.01$ & 
$-0.03$  & $-0.02$ & $+0.07$ & $-0.01$ \\
\ion{Zr}{2} & $+$ & $-0.03$ & $0$ & $-0.04$ & $0$ & 
$+0.06$  & $+0.07$ & $-0.03$ & $+0.03$ & 
$+0.01$  & $+0.03$ & $-0.07$ & $0$ \\
           & $-$ & $+0.04$ & $+0.01$ & $+0.05$ & $+0.01$ & 
$-0.05$  & $-0.06$ & $+0.03$ & $-0.01$ & 
$-0.01$  & $+0.01$ & $+0.09$ & $0$ \\
\ion{Mo}{1} & $+$ & ${\bf +0.30}$ & $-0.08$ & $-0.01$ & $-0.05$ & 
${\bf +0.23}$  & $-0.02$ & $-0.02$ & $-0.01$ & 
${\bf +0.31}$  & $-0.05$ & $-0.02$ & $-0.03$ \\
           & $-$ & ${\bf -0.37}$ & $+0.09$ & $+0.02$ & $+0.05$ & 
${\bf -0.25}$  & $+0.03$ & $+0.02$ & $+0.01$ & 
${\bf -0.35}$  & $+0.07$ & $+0.03$ & $+0.03$ \\
\ion{Ru}{1} & $+$ & ${\bf +0.36}$ & $-0.08$ & $-0.02$ & $-0.05$ & 
${\bf +0.24}$  & $-0.03$ & $-0.01$ & $-0.01$ & 
${\bf +0.32}$  & $-0.05$ & $-0.02$ & $-0.03$ \\
           & $-$ & ${\bf -0.39}$ & $+0.09$ & $+0.02$ & $+0.06$ & 
${\bf -0.26}$  & $+0.03$ & $+0.02$ & $+0.01$ & 
${\bf -0.33}$  & $+0.08$ & $+0.02$ & $+0.03$ \\
\ion{Rh}{1} & $+$ & ${\bf +0.31}$ & $-0.08$ & $-0.01$ & $-0.05$ & 
${\bf +0.24}$  & $-0.02$ & $-0.01$ & $-0.01$ & 
${\bf +0.31}$  & $-0.05$ & $0$ & $-0.03$ \\
           & $-$ & ${\bf -0.35}$ & $+0.09$ & $+0.01$ & $+0.05$ & 
${\bf -0.25}$  & $+0.03$ & $+0.01$ & $+0.01$ & 
${\bf -0.33}$  & $+0.06$ & $+0.01$ & $+0.02$ \\
\ion{Pd}{1} & $+$ & ${\bf +0.28}$ & ${\bf -0.11}$ & $-0.02$ & $-0.06$ & 
${\bf +0.24}$  & $-0.03$ & $-0.02$ & $-0.01$ & 
${\bf +0.28}$  & $-0.06$ & $-0.02$ & $-0.04$ \\
           & $-$ & ${\bf -0.29}$ & ${\bf +0.12}$ & $+0.02$ & $+0.06$ & 
${\bf -0.25}$  & $+0.04$ & $+0.02$ & $+0.01$ & 
${\bf -0.27}$  & $+0.10$ & $+0.02$ & $+0.03$ \\
\ion{Ag}{1} & $+$ & ${\bf +0.31}$ & $-0.10$ & $-0.01$ & $-0.06$ & 
${\bf +0.24}$  & $-0.03$ & $-0.01$ & $-0.01$ & 
\nodata  & \nodata  & \nodata  & \nodata  \\
           & $-$ & ${\bf -0.34}$ & $+0.10$ & $+0.01$ & $+0.06$ & 
${\bf -0.25}$  & $+0.04$ & $+0.01$ & $+0.01$ & 
\nodata  & \nodata  & \nodata  & \nodata  \\
\ion{Ba}{2} & $+$ & $+0.08$ & $+0.06$ & $-0.10$ & $+0.02$ & 
$+0.10$  & $+0.06$ & ${\bf -0.13}$ & $+0.02$ & 
$+0.05$  & $+0.08$ & ${\bf -0.11}$ & $+0.03$ \\
           & $-$ & $-0.06$ & $-0.05$ & ${\bf +0.12}$ & $-0.01$ & 
$-0.10$  & $-0.06$ & ${\bf +0.14}$ & $-0.01$ & 
$-0.05$  & $-0.06$ & ${\bf +0.12}$ & $-0.02$ \\
\ion{La}{2} & $+$ & $+0.01$ & $+0.02$ & $-0.03$ & $+0.01$ & 
$+0.08$  & $+0.07$ & $-0.04$ & $+0.03$ & 
$+0.05$  & $+0.06$ & $-0.02$ & $+0.02$ \\
           & $-$ & $0$ & $0$ & $+0.03$ & $0$ & 
$-0.07$  & $-0.06$ & $+0.06$ & $-0.02$ & 
$-0.04$  & $-0.05$ & $+0.02$ & $-0.02$ \\
\ion{Ce}{2} & $+$ & $+0.03$ & $+0.02$ & $-0.01$ & $+0.01$ & 
$+0.08$  & $+0.07$ & $-0.02$ & $+0.03$ & 
$+0.04$  & $+0.05$ & $-0.01$ & $+0.02$ \\
           & $-$ & $-0.01$ & $-0.01$ & $+0.01$ & $+0.01$ & 
$-0.07$  & $-0.05$ & $+0.02$ & $-0.02$ & 
$-0.04$  & $-0.05$ & $+0.01$ & $-0.02$ \\
\ion{Pr}{2} & $+$ & $+0.05$ & $+0.03$ & $-0.02$ & $+0.01$ & 
$+0.10$  & $+0.07$ & $-0.02$ & $+0.03$ & 
$+0.06$  & $+0.06$ & $-0.02$ & $+0.02$ \\
           & $-$ & $-0.04$ & $-0.02$ & $+0.02$ & $0$ & 
$-0.09$  & $-0.05$ & $+0.03$ & $-0.02$ & 
$-0.05$  & $-0.05$ & $+0.02$ & $-0.02$ \\
\ion{Nd}{2} & $+$ & $+0.04$ & $+0.02$ & $-0.02$ & $+0.01$ & 
$+0.09$  & $+0.07$ & $-0.03$ & $+0.03$ & 
$+0.06$  & $+0.06$ & $-0.01$ & $+0.02$ \\
           & $-$ & $-0.02$ & $-0.01$ & $+0.03$ & $0$ & 
$-0.08$  & $-0.05$ & $+0.04$ & $-0.02$ & 
$-0.05$  & $-0.05$ & $+0.01$ & $-0.02$ \\
\ion{Sm}{2} & $+$ & $+0.03$ & $+0.02$ & $-0.01$ & $+0.01$ & 
$+0.09$  & $+0.08$ & $-0.01$ & $+0.03$ & 
$+0.05$  & $+0.05$ & $0$ & $+0.02$ \\
           & $-$ & $-0.01$ & $-0.01$ & $+0.01$ & $+0.01$ & 
$-0.07$  & $-0.06$ & $+0.01$ & $-0.02$ & 
$-0.04$  & $-0.05$ & $+0.01$ & $-0.02$ \\
\ion{Eu}{2} & $+$ & $0$ & $+0.01$ & $-0.06$ & $0$ & 
$+0.07$  & $+0.06$ & $-0.09$ & $+0.02$ & 
$+0.03$  & $+0.05$ & $-0.04$ & $+0.01$ \\
           & $-$ & $0$ & $0$ & $+0.07$ & $+0.01$ & 
$-0.07$  & $-0.06$ & ${\bf +0.12}$ & $-0.01$ & 
$-0.03$  & $-0.02$ & $+0.05$ & $-0.01$ \\
\ion{Gd}{2} & $+$ & $-0.01$ & $-0.01$ & $-0.03$ & $0$ & 
$+0.08$  & $+0.07$ & $-0.04$ & $+0.03$ & 
$+0.04$  & $+0.04$ & $-0.02$ & $+0.01$ \\
           & $-$ & $+0.01$ & $+0.03$ & $+0.03$ & $+0.01$ & 
$-0.07$  & $-0.05$ & $+0.06$ & $-0.01$ & 
$-0.04$  & $-0.01$ & $+0.02$ & $-0.01$ \\
\ion{Tb}{2} & $+$ & $-0.02$ & $-0.01$ & $-0.02$ & $-0.01$ & 
$+0.09$  & $+0.07$ & $-0.02$ & $+0.03$ & 
$+0.05$  & $+0.04$ & $-0.01$ & $+0.01$ \\
           & $-$ & $+0.03$ & $+0.04$ & $+0.03$ & $+0.01$ & 
$-0.08$  & $-0.05$ & $+0.02$ & $-0.01$ & 
$-0.04$  & $-0.02$ & $+0.01$ & $-0.01$ \\
\ion{Dy}{2} & $+$ & $-0.01$ & $0$ & $-0.04$ & $0$ & 
$+0.08$  & $+0.07$ & $-0.05$ & $+0.03$ & 
$+0.05$  & $+0.05$ & $-0.01$ & $+0.02$ \\
           & $-$ & $+0.02$ & $+0.02$ & $+0.05$ & $+0.01$ & 
$-0.07$  & $-0.05$ & $+0.07$ & $-0.01$ & 
$-0.04$  & $-0.04$ & $+0.02$ & $-0.02$ \\
\ion{Ho}{2} & $+$ & $-0.06$ & $-0.03$ & $-0.05$ & $-0.01$ & 
$+0.08$  & $+0.06$ & $-0.07$ & $+0.02$ & 
$+0.04$  & $+0.03$ & $-0.02$ & $+0.01$ \\
           & $-$ & $+0.04$ & $+0.05$ & $+0.06$ & $+0.02$ & 
$-0.07$  & $-0.04$ & $+0.10$ & $-0.01$ & 
$-0.04$  & $-0.01$ & $+0.02$ & $-0.01$ \\
\pagebreak
\ion{Er}{2} & $+$ & $-0.04$ & $-0.03$ & $-0.05$ & $-0.01$ & 
$+0.08$  & $+0.06$ & $-0.09$ & $+0.02$ & 
$+0.04$  & $+0.03$ & $-0.02$ & $+0.01$ \\
           & $-$ & $+0.03$ & $+0.05$ & $+0.06$ & $+0.02$ & 
$-0.08$  & $-0.05$ & ${\bf +0.12}$ & $-0.01$ & 
$-0.03$  & $-0.01$ & $+0.03$ & $-0.01$ \\
\ion{Tm}{2} & $+$ & $-0.02$ & $-0.01$ & $-0.01$ & $0$ & 
$+0.08$  & $+0.07$ & $-0.02$ & $+0.03$ & 
$+0.04$  & $+0.04$ & $-0.01$ & $+0.01$ \\
           & $-$ & $+0.03$ & $+0.03$ & $+0.02$ & $+0.01$ & 
$-0.07$  & $-0.05$ & $+0.02$ & $-0.01$ & 
$-0.03$  & $-0.02$ & $+0.01$ & $-0.01$ \\
\ion{Yb}{2} & $+$ & $-0.02$ & $-0.03$ & ${\bf -0.11}$ & $-0.02$ & 
$+0.07$  & $+0.06$ & ${\bf -0.18}$ & $+0.01$ & 
$+0.01$  & $0$ & $-0.12$ & $-0.01$ \\
           & $-$ & $-0.01$ & $+0.04$ & ${\bf +0.13}$ & $+0.02$ & 
$-0.06$  & $-0.04$ & ${\bf +0.26}$ & $-0.01$ & 
$-0.01$  & $+0.07$ & ${\bf +0.17}$ & $+0.02$ \\
\ion{Lu}{2} & $+$ & \nodata & \nodata & \nodata & \nodata & 
$+0.05$  & $+0.08$ & $0$ & $+0.03$ & 
$+0.03$  & $+0.01$ & $-0.03$ & $0$ \\
           & $-$ & \nodata & \nodata & \nodata & \nodata & 
$-0.03$  & $-0.06$ & $+0.01$ & $-0.02$ & 
$-0.03$  & $+0.04$ & $+0.03$ & $0$ \\
\ion{Hf}{2} & $+$ & $+0.01$ & $+0.01$ & $-0.01$ & $0$ & 
$+0.08$  & $+0.07$ & $-0.01$ & $+0.03$ & 
$+0.04$  & $+0.05$ & $0$ & $+0.02$ \\
           & $-$ & $+0.01$ & $0$ & $+0.01$ & $0$ & 
$-0.07$  & $-0.05$ & $+0.01$ & $-0.02$ & 
$-0.03$  & $-0.05$ & $0$ & $-0.01$ \\
\ion{Os}{1} & $+$ & ${\bf +0.29}$ & $-0.03$ & $-0.01$ & $-0.02$ & 
${\bf +0.26}$  & $0$ & $-0.03$ & $02$ & 
${\bf +0.26}$  & $0$ & $0$ & $0$ \\
           & $-$ & ${\bf -0.24}$ & $+0.04$ & $+0.01$ & $+0.02$ & 
${\bf -0.26}$  & $+0.01$ & $+0.03$ & $0$ & 
${\bf -0.21}$  & $0$ & $0$ & $0$ \\
\ion{Ir}{2} & $+$ & ${\bf +0.14}$ & $-0.03$ & $-0.02$ & $-0.02$ & 
${\bf +0.22}$  & $+0.02$ & $-0.02$ & $+0.01$ & 
\nodata  & \nodata & \nodata & \nodata \\
           & $-$ & $-0.06$ & $+0.06$ & $+0.03$ & $+0.02$ & 
${\bf -0.20}$  & $0$ & $+0.02$ & $0$ & 
\nodata  & \nodata & \nodata & \nodata \\
\ion{Th}{2} & $+$ & $+0.04$ & $+0.01$ & $-0.01$ & $0$ & 
$+0.10$  & $+0.07$ & $-0.01$ & $+0.03$ & 
\nodata  & \nodata & \nodata & \nodata \\
           & $-$ & $-0.02$ & $0$ & $+0.01$ & $0$ & 
$-0.09$  & $-0.05$ & $+0.01$ & $-0.02$ & 
\nodata  & \nodata & \nodata & \nodata \\
\enddata
\tablenotetext{a}{The top number indicates the abundance offset when the parameter is increased by the given amount, while the bottom number indicates the abundance offset when the parameter is decreased.  Abundance offsets larger than 0.1 dex are bolded.}
\end{deluxetable*}

\subsection{Discussion of Certain Spectral Features}\label{sub:lines}
\begin{figure*}[h!]
\centering
\hspace*{-0.25in}
    \includegraphics[scale=0.75,trim=0.in 0.1in 0in 0.0in,clip]{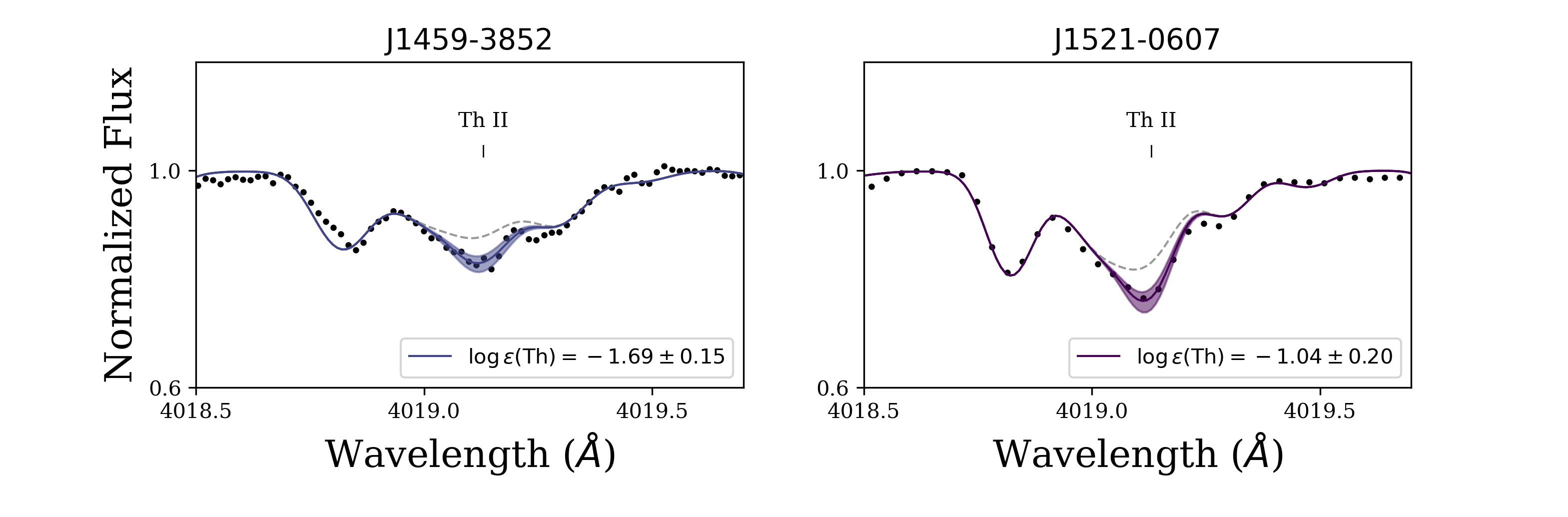}
    \caption{Example syntheses of the \ion{Th}{2} line in J1459$-$3852 (left) and J1521$-$0607 (right).  The black points show the spectrum, the solid lines show the best-fit syntheses, and the shaded regions show abundance uncertainties.  The dashed gray lines show the synthesis with zero abundance for the line of interest.}
    \label{fig:ThSynths}
\end{figure*}

In general, lines were removed from analyses if they were prohibitively strong, weak, or blended.  As an example, the strong 4554~{\AA} \ion{Ba}{2} line was excluded from all three stars, while the 4215~{\AA} \ion{Sr}{2} line was excluded from two stars.  Many of the elements considered in this analysis have many available spectral lines, which means that only the best lines could be selected.  However, some important elements have only a few lines.  These elements are discussed below.
\begin{description}
    \item[Pd and Ag] There \cms{are three \ion{Pd}{1} lines at 3404.58~{\AA}, 3460.74 \AA, and 3516.94 \AA.  There is a single \ion{Ag}{1} line at 3382.89~{\AA}.  The second \ion{Pd}{1} line is not detectable in J1459$-$3852, though all three were detectable in J1521$-$0607 and J1944$-$4039.}  The Ag line was not detected in J1944$-$4039, nor did it yield a meaningful upper limit.\\
    
    \item[Yb] The line at 3694.19~{\AA} was detected in all three stars.  As with other lines, the $r$-process isotopic ratios of \citet{2008ARA&A..46..241S} were adopted.\\
    
    \item[Pb] The Pb line at 4057.807~{\AA} was not detected in any of the three stars.  It is blended with  a strong Mg line and is too weak to be seen, especially in the more metal-rich stars.  Upper limits are provided in Table \ref{table:allabundances}.\\
    
    \item[Th] There are three potential \ion{Th}{2} lines, at 4019.13, 4086.52, and 4094.75~{\AA}.  Of the three, 4019.13~{\AA} is the strongest, but it is blended with \ion{Fe}{1} and \ion{Co}{1}. The blends mean that only an upper limit is possible in J1944$-$4039.  The 4086.52~{\AA} line is next to the \ion{La}{2} feature in Figure \ref{fig:Synths}, and is barely visible.  The 4094.75~{\AA} line is blended with \ion{Er}{2} and CH.  For these reasons, only 4019.13~{\AA} is included.  Syntheses of the 4019.13~{\AA} line in J1459$-$3852 and J1521$-$0607 are shown in Figure \ref{fig:ThSynths}.\\
    
    \item[U] There are three potential \ion{U}{2} lines in the observed spectral region \citep{2023ApJ...948..122S}.  The line at 3859.57~{\AA} is in the wing of a fairly strong \ion{Fe}{1} line and can be severely blended with CN and \ion{Nd}{2} lines, providing only an upper limit in J1459$-$3852.  The 4050.04~{\AA} line is blended with \ion{La}{2} and does not fit well in any of the three stars.  Finally, the 4090.13~{\AA} line is blended with an \ion{Fe}{1} line, but provides a more constraining upper limit for J1459$-$3852.
    \end{description}

\section{Abundances}\label{sec:results}
\subsection{Mean Abundances and Classifications}\label{sub:class}
The final mean $\log \epsilon(X)$ abundances, uncertainties, and the number of spectral lines for each element are given in Table \ref{table:allabundances}, as well as the abundance ratios ([Eu/Fe], [Ba/Eu], and [Sr/Ba]) that are used in classifying the stars.  These abundances produce [Eu/Fe] ratios that place J1459$-$3852 and J1521$-$0607 in the $r$-II category ($[\rm{Eu/Fe}]>+0.7$ and $[\rm{Ba/Eu}]<0$), in agreement with the original classifications from \cite{2020ApJS..249...30H} and \cite{2018ApJ...868..110S}, respectively.  J1944$-$4039 is found to fall just below the $r$-I definition, in contrast to \citet[although the disagreement in Eu is not significant, as discussed below]{2020ApJS..249...30H}.

The low and subsolar [Ba/Eu] ratios show that J1459$-$3852 and J1521$-$0607 are consistent with showing minimal signs of $s$-process contamination.  For reference, the solar $r$-process ratio is $[\rm{Ba/Eu}] = -0.9$ \citep{2000ApJ...544..302B}. The higher [Ba/Eu] ratio of J1944$-$4039 could indicate some contributions from the $s$-process, although the subsolar ratio is still consistent with a dominant $r$-process.  The [Sr/Ba] ratios for all three stars rule out a limited-$r$ signature, which refers to enhancements in the lighter $r$-process elements, compared to the main $r$-process.

\subsection{Comparisons with the Literature}\label{sub:litcomp}
\begin{figure}[h!]
\begin{center}
\centering\hspace*{-0.45in}
\includegraphics[scale=0.62,trim=0.0in 0.in 0.in 0.in,clip]{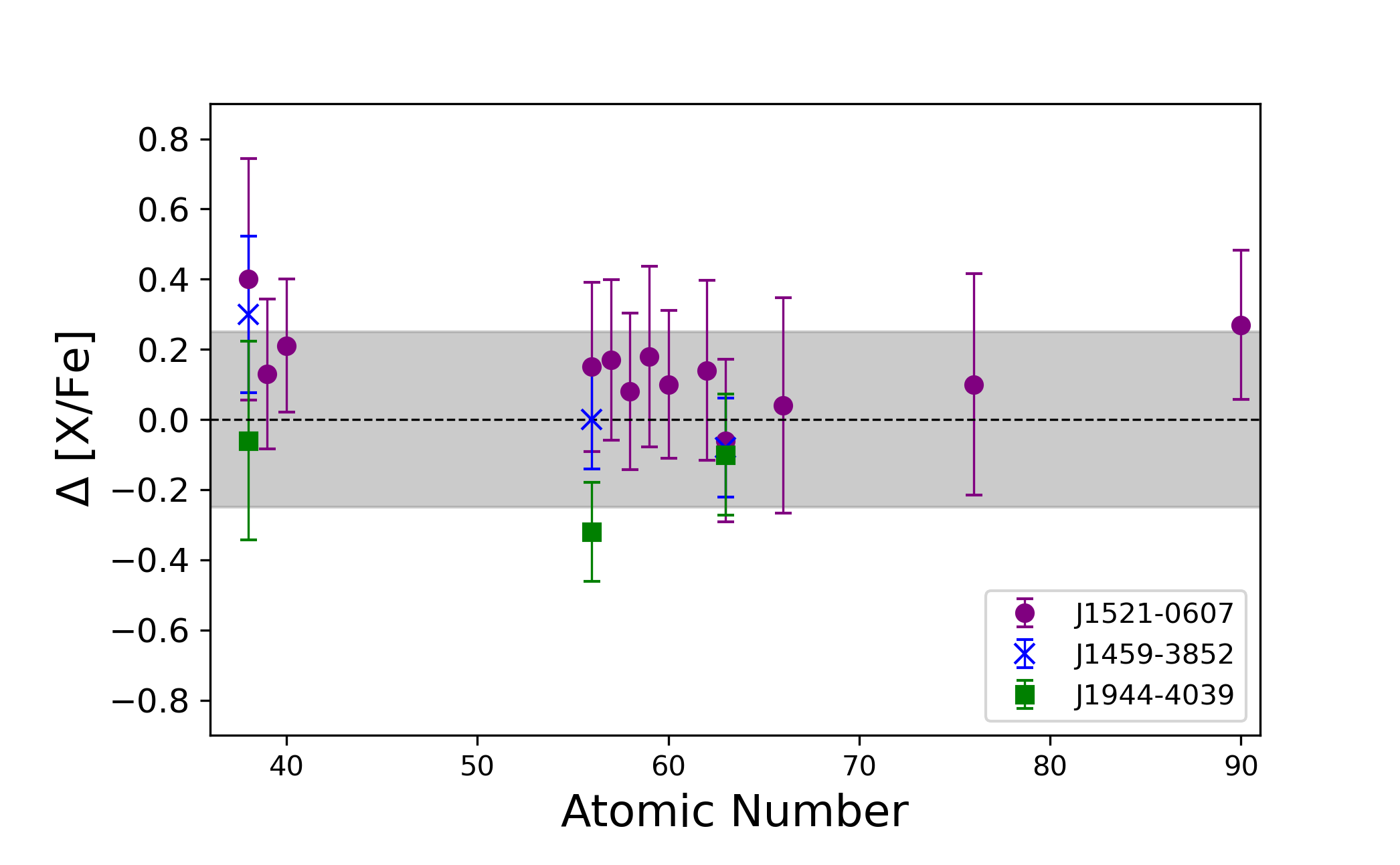}
\caption{Differences between abundances measured in this portrait analysis compared to the snapshot analyses from the literature (\citealt{2018ApJ...868..110S} for J1521$-$0607; \citealt{2020ApJS..249...30H} for J1459$-$3852 and J1944$-$4039). The dashed horizontal line shows equal agreement, while the light gray band shows offsets of $\pm0.25$ dex.}   \label{fig:literatureplot}
\end{center}
\end{figure}

All three stars have previously been analyzed in lower-resolution snapshot analyses. Figure \ref{fig:literatureplot} shows the derived neutron-capture [X/Fe] abundance ratios of the targets compared to the abundances in the literature from \cite{2018ApJ...868..110S} for J1521$-$0607 and from \cite{2020ApJS..249...30H} for J1459$-$3852 and J1944$-$4039. The uncertainties in the abundance offsets include uncertainties from this paper and those from the literature, added in quadrature.\footnote{Uncertainties were not included in \cite{2020ApJS..249...30H}, so representative uncertainties of 0.1 dex were assumed for Ba and Eu, while 0.2 dex was adopted for Sr.} \citet{2020ApJS..249...30H} only published Sr, Ba, and Eu abundances, while \citet{2018ApJ...868..110S} published a wider variety of elements, including Th.

Figure \ref{fig:literatureplot} shows good agreement within the uncertainties for most elements, with a few exceptions.  This paper finds a lower Ba abundance for J1944$-$4039, compared to \cite{2020ApJS..249...30H}. 
For J1459$-$3852 and J1521$-$0607, Sr is found to be higher in this work, compared to \citet{2020ApJS..249...30H} and \citet{2018ApJ...868..110S}.  The offset for J1521$-$0607 is due to a typo in \cite{2018ApJ...868..110S}---the intended value of $[{\rm Sr/Fe}] = +0.52\pm0.20$ would put the two in agreement within their $1\sigma$ uncertainties. Th is also found to be higher in J1521$-$0607, compared to \citet{2018ApJ...868..110S}, which may be due to the blends that are better modeled in a higher-resolution portrait spectrum.  The different atmospheric parameters between this analysis and the snapshot analyses will also lead to small offsets in abundances.  \cms{The 2K parameters lead to small differences from the snapshot parameters: $|\Delta T_{\rm{eff}}|<150$\;K, $|\Delta \log g |<0.3$\,dex, $|\Delta \xi|<0.2$\; km s$^{-1}$, and $|\Delta [\rm{Fe/H}]|<0.25$\,dex.}

%Lit offset figure

\section{Discussion}\label{sec:discussion}
The neutron-capture abundance patterns are compared to other metal-poor stars and the Sun in Section \ref{sub:ncapcompare}, and are used to determine cosmo-chronometric ages in Section \ref{sub:actinides}.  Finally, the abundances are discussed in the context of Galactic Archaeology, including potential links to the Thamnos substructure, in Section \ref{sub:implications}.

%abundance plots
\begin{figure*}[ht!]
\begin{center}
\centering
\hspace*{-0.45in}
\subfigure{\includegraphics[scale=0.78,trim=0 0.45in 0.45in 0.3in,clip]{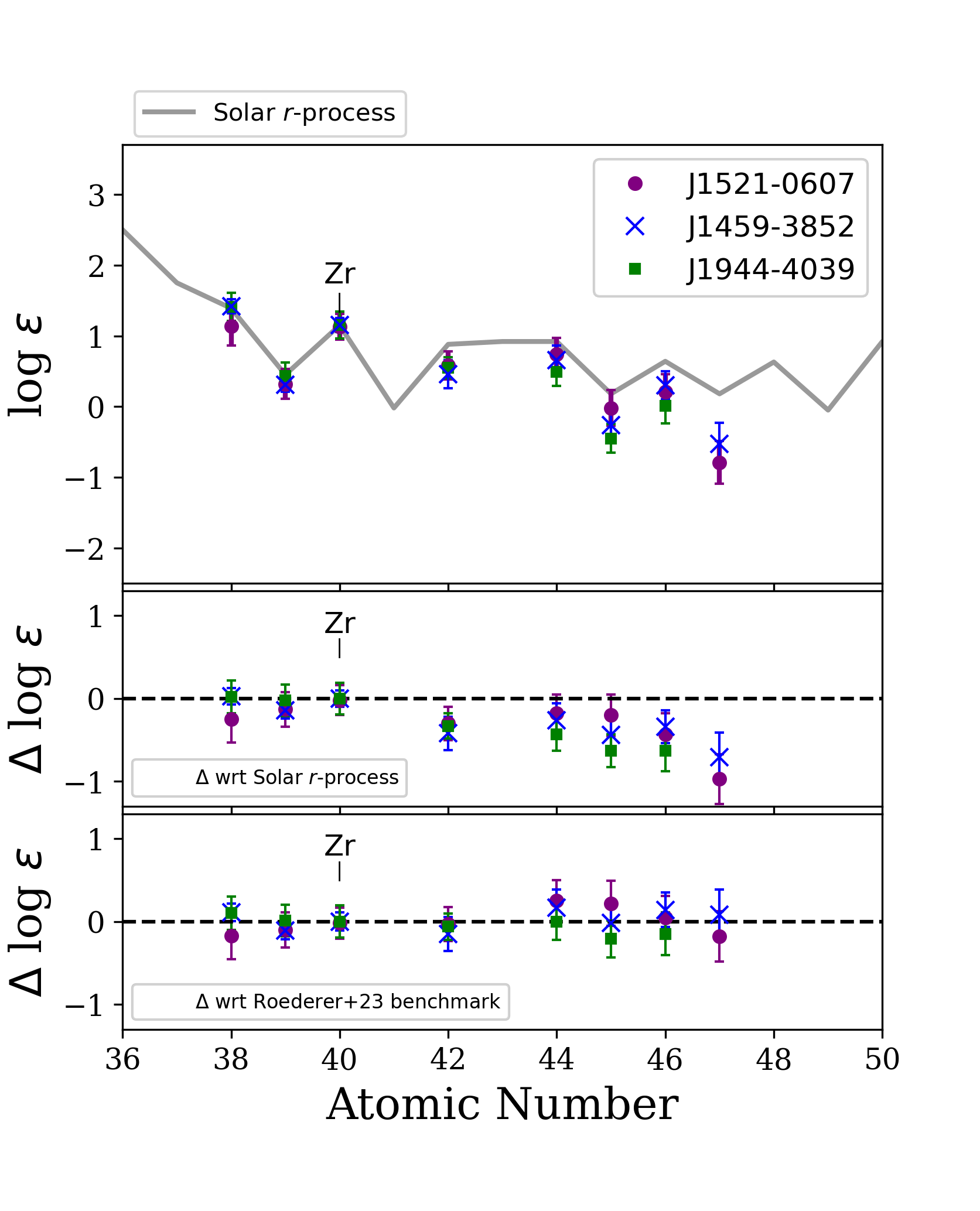}\label{subfig:lightr}}
\subfigure{\includegraphics[scale=0.78,trim=0.67in 0.45in 0.55in 0.3in,clip]{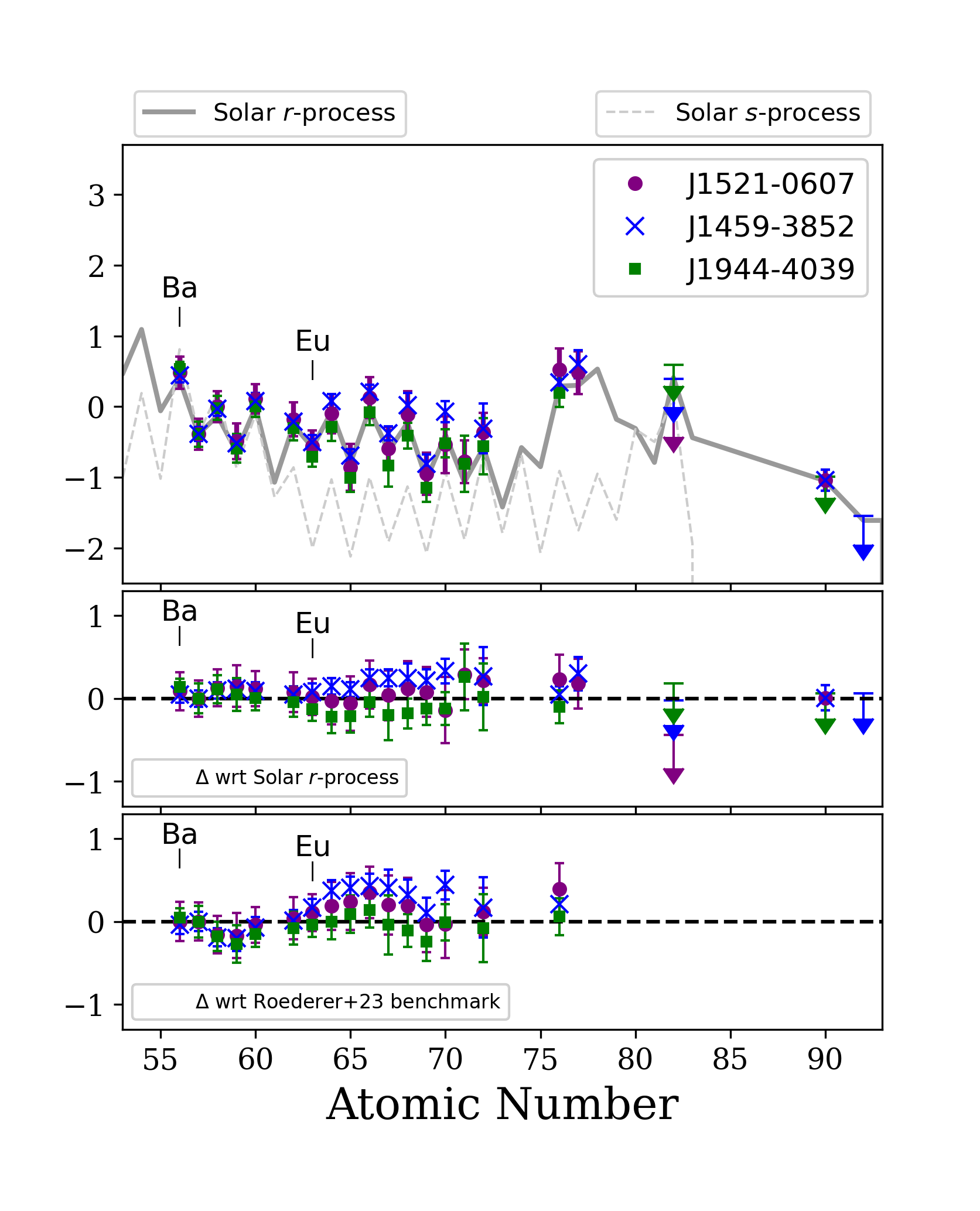}\label{subfig:heavyr}}
\caption{Abundance patterns of the three target stars with respect to (wrt) to various benchmark patterns for the light $r$-process elements ($Z\leq 47$; left) and the heavy $r$-process elements ($Z\geq 56$; right).  The light $r$-process patterns are scaled to the same Zr abundance, while the heavy $r$-process patterns are scaled to the same La abundance.  The top panels show the full $r$-process abundance patterns of the three target stars compared to the Solar $s$- and $r$-process patterns from \citet{1999ApJ...525..886A}, while the middle panel shows the residual offsets from the Solar $r$-process pattern.  The bottom panel shows the offsets relative to the metal-poor star benchmark pattern from \citet{2023Sci...382.1177R}.  The uncertainties in $\Delta \log \epsilon(X)$ reflect uncertainties from this paper and the benchmark values, added in quadrature.}
    \label{fig:thesisabundanceplot}
\end{center}
\end{figure*}

\subsection{Patterns of the Neutron-capture Elements}\label{sub:ncapcompare}
Figure \ref{fig:thesisabundanceplot} presents the relative abundances of the neutron-capture elements.  The abundances are divided into two groups: 1) the lighter $r$-process elements (Sr through Ag), as shown in the left panel, and the heavier $r$-process elements (Ba through U), as shown in the right panel.  The abundances of the three target stars are plotted along with the Solar $s$-process pattern (dashed light gray) and the (presumed) $r$-process residual (Solar gray) from \citet{1999ApJ...525..886A}.  To examine the relative abundance patterns, the abundances are all normalized to a common Zr abundance in the left panel and a common La abundance in the right panel (see the discussion below).  The bottom two panels show residual plots with regard to the Solar $r$-process pattern (middle) and the metal-poor star ``baseline'' pattern from \citet[bottom]{2023Sci...382.1177R}, which was constructed from 13 metal-poor stars with [Eu/Fe] $< +0.3$.\\

\subsubsection{Patterns of the Lighter Neutron-capture Elements}\label{subsub:lightr}
The left panel of Figure \ref{fig:thesisabundanceplot} shows a common trend for metal-poor stars: compared to the Solar pattern, Ru, Rh, Pd, and Ag are significantly lower in all three stars.  \cms{The lower left panel compares the abundance patterns to the benchmark pattern from \citet{2023Sci...382.1177R}, which is based on abundances from 13 stars that are not $r$-process-enhanced (i.e., that have $[\rm{Eu/Fe}] < +0.3$).  For these lighter $r$-process elements, all three targets are consistent with this benchmark pattern.}

A previous analysis by \citet{2022ApJ...936...84R} showed that elements with $38\leq Z \leq 42$ (including Sr, Y, and Zr) exhibit a consistent, ``universal'' pattern, while the pattern of elements with $44\leq Z \leq 50$ (including Ru, Rh, Pd, and Ag) varies between metal-poor stars.  \citet{2023Sci...382.1177R} interpret this variation as resulting from differing amounts of fission fragments from transuranic elements synthesized by the $r$-process.  They show that the abundances of these elements, compared to Zr, increase with [Eu/Fe], suggesting that stars with more enhancement in the $r$-process should have more relative enhancement in fission fragments.  Indeed, Figure \ref{fig:thesisabundanceplot} hints at this trend: J1459$-$3852 and J1521$-$0607, the two $r$-II stars in this sample, exhibit higher [Ru/Zr], [Rh/Zr], and [Pd/Zr] ratios than J1944$-$4039, which is not $r$-process enhanced. Unfortunately, the upper limit on Ag in J1944$-$4039 is not useful for testing if Ag is also lower than the benchmark value.  

\cms{J1521$-$0607 and J1944$-$4039 both have low $\log \epsilon(\rm{Pd/Zr})$ ratios than might be expected for their [Eu/Fe] ratios (see, e.g., \citealt{2024ApJ...971..158R}).  However, these low values are consistent with the observed ranges in the benchmark sample.}

\subsubsection{The Patterns of the Heavier Neutron-capture Elements}\label{sub:heavyr}
The right panel of Figure \ref{fig:thesisabundanceplot} shows the patterns of the heavier elements, scaled to a common La abundance.  \citet{2023Sci...382.1177R} found that the star-to-star dispersion among the lighter lanthanides with $56\leq Z \leq 62$ (Ba to Sm in this sample) was negligible, indicating a robust pattern for these elements.  The abundances of the heavier $r$-process elements exhibit the well-known trend of metal-poor stars, where the abundances of the lanthanides, Os, and Ir, are generally consistent with the Solar $r$-process pattern (e.g., \citealt{2008ARA&A..46..241S}).

\citet{2023Sci...382.1177R} also found that stars with higher $r$-process enrichment tend to have slightly higher abundances of elements from $63\leq Z \leq 78$ (Eu to Ir in this sample).  This is evident \cms{for the target stars} in the bottom right panel of Figure \ref{fig:thesisabundanceplot}, which compares the target stars with the benchmark metal-poor star pattern.  \cms{As expected for a non-$r$-process-enhanced metal-poor star,} J1944$-$4039 has a pattern that is consistent with the benchmark star, while the $r$-II stars J1459$-$3852 and J1521$-$0607 tend to show higher abundances of Eu to Ir.  Similar to the light elements, \citet{2023Sci...382.1177R} suggest that this is a signature of fission fragments from transuranic elements, where stars with more $r$-process-enhancement have larger contributions from these fragments.

\subsection{The Actinides: Thorium and Uranium}\label{sub:actinides}
Th and U can only form in the $r$-process, and they are both radioactive, with long and well-known half-lives (14.0 Gyr for Th and  4.5 Gyr for U; e.g., \citealt{Holden1990,Schon2004}). Given this, the age of the $r$-process material can be calculated by comparing the abundances of Th and U to stable $r$-process elements, assuming some initial production ratios (PR) and using equations 1-3 from \citet{2017ApJ...844...18P}.  

\begin{figure}[ht!]
\begin{center}
\centering
%\hspace*{-0.25in}
%\subfigure{\includegraphics[scale=0.6,trim=0 0.1in 0.45in 0.15in,clip]{AgesHill.png}\label{subfig:AgesHill}}
\hspace*{-0.25in}
%\subfigure{
\includegraphics[scale=0.58,trim=0.11in 0.05in 0.68in 0.45in,clip]{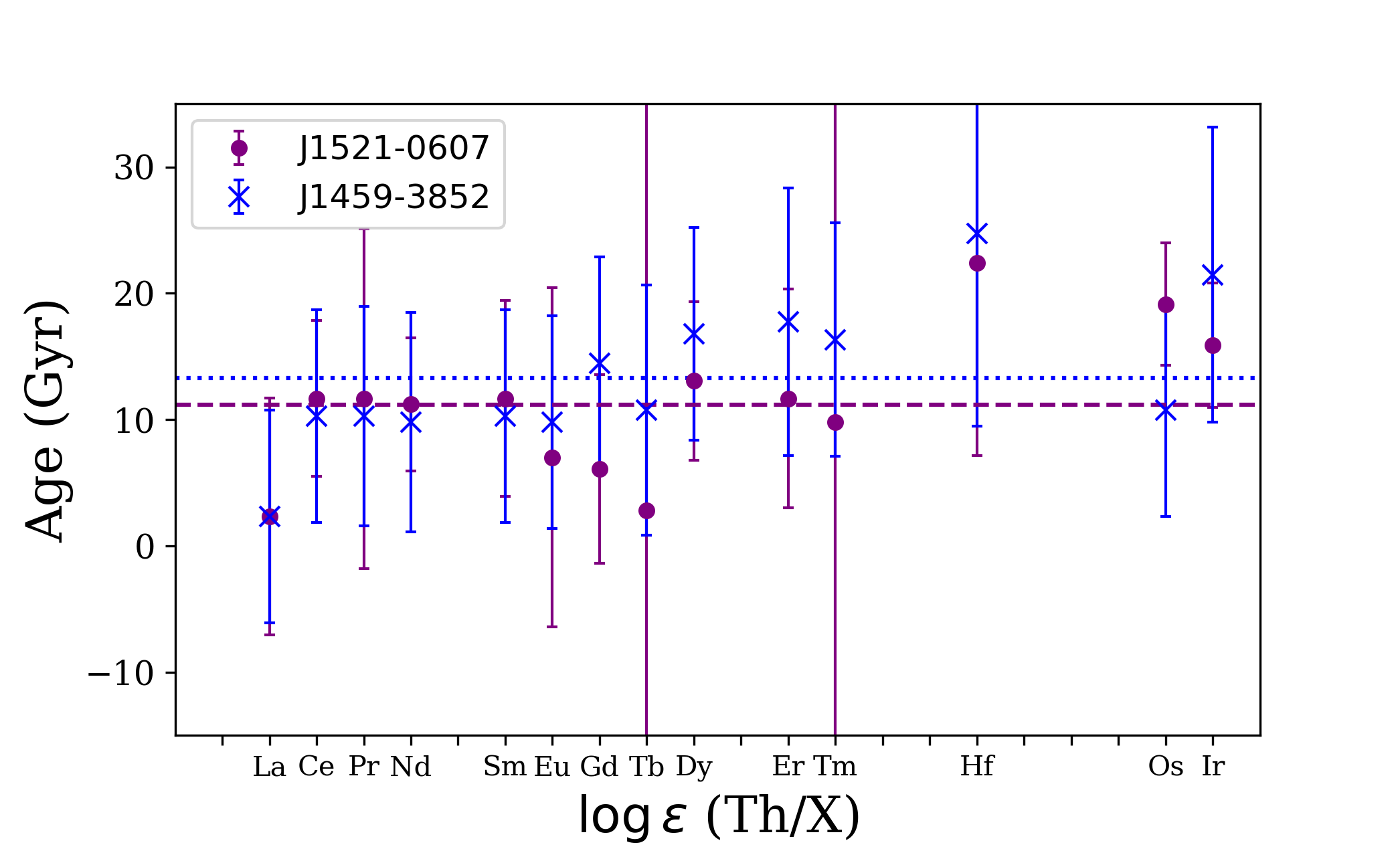}\label{subfig:AgesSchatz}
\caption{Ages determined from $\log \epsilon(\rm{Th/X})$ abundance ratios, using the production ratios from \citet{2002ApJ...579..626S} and the equations given in \citet{2017ApJ...844...18P}.  The uncertainties represent the uncertainties in the derived $\log \epsilon(\rm{Th/X})$ ratios and the PRs.  The horizontal dashed and dotted lines show the average ages for J1521$-$0607 and J1459$-$3852, respectively.  Both stars are consistent with old ages for the $r$-process material.}
    \label{fig:Ages}
\end{center}
\end{figure}

Th was measured in both J1459$-$3852 and J1521$-$0607 and an upper limit for U was determined for J1459$-$3852.  Figure \ref{fig:Ages} shows ages calculated with $\log \epsilon(\rm{Th/X})$ ratios, using the PRs determined from waiting-point calculations by \citet{2002ApJ...579..626S}.  The uncertainties reflect uncertainties in the $\log \epsilon(\rm{Th/X})$ ratios and the uncertainties in the PR calculations quoted in \citet{2002ApJ...579..626S}.  The ages determined from the $\log \epsilon(\rm{Th/X})$ ratios are fairly consistent between the elements.\footnote{PRs are also available from \citet{2017Hill}, based on a high-entropy wind model.  These ages are not shown, but they produce slightly older average ages, with greater dispersion between the different ratios.}  The average age for J1459$-$3852 is $13.3\pm5.5$ Gyr, while the average age is $11.2\pm5.4$ Gyr for J1521$-$0607.  The more reliable chronometer, $\log \epsilon(\rm{Th/U})$, yields a lower limit for the age of $8.4$ Gyr.  Note that the upper limits in Th in J1944$-$4039 produce negative lower limits in age, which are not useful.

\cms{The $\log \epsilon(\rm{Th/Eu})$ ratios in J1459$-$3852 and J1521$-$0607 ($-0.59$ and $-0.55$) are both consistent with old ages, as expected for typical metal-poor stars} and do not indicate the presence of an actinide boost (e.g., \citealt{2018ApJ...859L..24H}, \citealt{2018ApJ...856..138J}, \citealt{placco2023}).

%abundance plots

\subsection{Connections with CDTGs and the Thamnos Stream}\label{sub:implications}

As discussed in Sections \ref{sec:Intro} \cms{and \ref{sub:assocCDTG}, the target stars in this paper were all previously found to have low-energy, retrograde orbits, supporting origins in low-mass, accreted systems,} to be part of CDTGs based on their clustering in dynamical space.  \cms{The chemical properties of the birth sites of these stars and potential links to CDTGs and streams are discussed below.}

\subsubsection{Properties of Birth Sites}

The low [Fe/H] ratio of J1459$-$3852 supports an origin in a low-mass system.   \citet{2023ApJ...946...48H} noted that their DTG-1 has a mean metallicity of $\langle [\rm{Fe/H}]\rangle = -2.78$.  The metallicity of J1459$-$3852 is consistent with an association with that DTG.  Its low [Fe/H]$=-$2.5
%, along with its enhanced [$\alpha$/Fe] ratio 
and old age (12~Gyr, see Section \ref{sub:actinides}), suggest that J1459$-$3852 formed in a low-mass galaxy with minimal chemical evolution, or that it was an early star to form in a galaxy that experienced subsequent chemical evolution.  The $r$-process enhancement suggests that it formed after early $r$-process nucleosynthesis.  Its retention of the $r$-process material suggests that it either formed in a low-mass system or that the $r$-process event produced a high amount of $r$-process elements (see, e.g., \citealt{2024ApJ...971..158R}).

The other two stars, J1521$-$0607 and \\ J1944$-$4039, have slightly higher metallicities, around $[\rm{Fe/H}]\sim-2$.  
%Both have enhanced [$\alpha$/Fe] ratios, indicating minimal contributions from Type Ia supernovae.  
While J1521$-$0607 is $r$-process enhanced, J1944$-$4039 is not.  Both stars have subsolar [Ba/Eu] ratios ($[\rm{Ba/Eu}] = -0.62$ for J1521$-$0607 and $[\rm{Ba/Eu}] = -0.42$ for J1944$-$4039), indicating minimal $s$-process contamination, which is further supported by the low Pb upper limits in both \citep{roederer10c}.  The disparate chemical abundances at the same [Fe/H] indicate that the two are likely from different birth sites.  J1521$-$0607 formed in an environment with significant $r$-process enhancement, while J1944$-$4039 either formed in an environment with minimal $r$-process nucleosynthesis, or it formed from prior $r$-process-enriched material that was diluted by subsequent stellar evolution.
%Fe nucleosynthesis, e.g., from Type II supernovae. 

%Note that \citet{2021ApJ...908...79G} grouped J1944$-$4039 in the same CDTG with J1459$-$3852.  The two stars could have a similar birth site if J1944$-$4039 formed after J1459$-$3852.  Compared to J1459$-$3852, J1944$-$4039 has $\Delta[\rm{Fe/H}] = +0.48$\,dex.  If the Fe enhancement in J1944$-$4039 was due to Type Ia supernova ejecta, one can infer the [Eu/Fe] it would have had without the supernova: $[\rm{Eu/Fe}]_{\rm{corr}} = 0.19+0.48 = +0.67$, which is slightly lower than J1459$-$3852.  This is somewhat similar to the trends seen in Ret~II (e.g., \citealt{2016ApJ...830...93J}).  However, it is more difficult to explain the different $r$-process patterns between the two stars.  As discussed in Section \ref{sub:ncapcompare}, J1944$-$4039 looks to be deficient in fission fragments relative to J1459$-$3852, indicating separate sources of $r$-process material.

\cms{
\subsubsection{Connections with CDTGs}
As shown in Table \ref{table:info}, all three stars have previously been associated with CDTGs or DTGs.}
Although the exact dynamical groups of stars differ in the above papers, J1521$-$0607 is consistently found to be in a dynamical grouping, while J1459$-$3852 and J1944$-$4039 are sometimes found to be in groups, depending on the stellar sample and method used for the clustering analysis. %CDTGs are an important part of Galactic archaeology because they connect stars through kinematics and chemical composition to find substructure and remnants of accretion.  

\cms{Various clustering analyses from the literature have placed the stars in different CDTGs, or none at all, depending on the sample of stars.}  With a sample of 446 r-process-enhanced metal-poor stars, \cite{2021ApJ...908...79G} placed both J1459$-$3852 and J1944$-$4039 in CDTG-2 and J1521$-$0607 in CDTG-27, though they argued that they two CDTGs could be associated with the same progenitor.  Subsequent analyses with different samples of stars found other results. From a sample of $\sim$1700 r-process-enhanced metal-poor stars, \citet{2023ApJ...943...23S} placed J1521$-$0607 in their CDTG-16,\footnote{Note that the numbers for each CDTG are automatically assigned in each paper, and have no connection across samples.} while J1459$-$3852 and J1944$-$4039 were not found to be associated with any CDTGs.  Based on 161 $r$-II stars, \citet{2023ApJ...946...48H} found that J1459$-$3852 was associated with their DTG-1, while J1521$-$0607 was associated with DTG-16 (note that J1944$-$4039 was not included in the Hattori et al. sample).  

Intriguingly, \citet{2023ApJ...946...48H} also include two other previously studied RPA stars in DTG-16, along with J1521$-$0607: J1538$-$1804, a strongly enhanced $r$-II star ($[\rm{Fe/H}]=-2.09$ and $[\rm{Eu/Fe}]=+1.27$; \citealt{2018ApJ...854L..20S}), and J0937$-$0607, a moderately enhanced $r$-II star with subsolar [$\alpha$/Fe] ratios ($[\rm{Fe/H}]=-1.86$, $[\rm{Eu/Fe}]=+0.85$, $[\rm{Mg/Fe}]=-0.25$; \citealt{2019ApJ...874..148S}).  Note that J0937$-$0607 was not included in the \citet{2021ApJ...908...79G} analysis, and it was not associated with a grouping in \citet{2023ApJ...943...23S}.  Neither \citet{2021ApJ...908...79G} nor \citet{2023ApJ...943...23S} found J1538$-$1804 to be associated with a CDTG. Based on the chemical abundances, specifically the differing [$\alpha$/Fe] ratios, it seems unlikely that J0937$-$0607 originated from the same environment as J1521$-$0607.  J1538$-$1804 also shows stronger $r$-process enrichment than J1521$-$0607.

Ultimately, the disparate abundance patterns among the three stars make it unlikely that all three are associated with the same birth site.  \cms{More work should be done to assess the chemical properties of stars in CDTGs and DTGs, to determine if they are real structures.}
%originated in the same birth site.  However, the chemical abundances of each star are consistent with origins in accreted low-mass systems.

\subsubsection{Connections to Thamnos}
\citet{2021ApJ...908...79G} suggested that all three stars could be connected to the Thamnos stream.  The Thamnos substructure was first identified in \citet{2019A&A...631L...9K}, who applied clustering methods to find large-scale structure in the halo. \citet{2019A&A...631L...9K} found that there was a grouping that stood out because of strong retrograde orbits, low metallicity, and $\alpha$-enhanced stars compared to other already-identified substructures. \cite{2019A&A...631L...9K} state that Thamnos likely originates from a dwarf galaxy with stellar mass $\lesssim$ 5$\times$10$^6 M_{\odot}$.   Since then, \citet{2023A&A...670A.106C} used Gaia DR3 spectra to determine [M/H], [Ca/Fe], and [Ce/Fe] in two Thamnos stars, finding $[\rm{M/H}]\sim -1.4$ and $-1.1$, with moderate Ca-enhancement and enhanced [Ce/Fe] ratios.  
\citet{2025A&A...698A.277D} determined the age-metallicity relations (AMRs) for Thamnos, finding that Thamnos is predominantly an old metal-poor stream, with half of the stars formed by $\sim 12.3\pm0.3$ Gyr, and metallicities ranging from $[\rm{Fe/H}] \sim -2.5$ to $-1.5$.

The findings of this paper are not inconsistent with a possible link between any of these three stars and Thamnos.
\begin{description}
\item[Metallicities] The three stars analyzed in this paper are all metal-poor ([Fe/H]$\leq-1.9$), within the metallicity range determined by \citet{2025A&A...698A.277D}.\\

\item [Ages] The old ages of J1459$-$3852 ($12 \pm 5$~Gyr) and J1521$-$0607 ($10 \pm 5$~Gyr) are consistent with the AMR of \citet{2025A&A...698A.277D}, especially given the large uncertainties in the abundances and the initial production ratios.\\

%\item[Light + $\alpha$-elements] All three stars are found to be $\alpha$-enhanced, with $[\alpha/\rm{Fe}]\approx +0.3$ \cms{CHECK THIS.}   These abundance ratios are consistent with the [Ca/Fe] ratios derived by \citet{2023A&A...670A.106C} for more metal-rich Thamnos stars.\\

\item[Neutron-capture elements] All three stars exhibit some $r$-process enrichment.  J1459$-$3852 and J1521$-$0607 are highly $r$-process enhanced.  \citet{2023A&A...670A.106C} found moderate [Ce/Fe] enhancement in two more metal-rich stars, using lower-resolution spectra, although those two stars could have $s$-process contamination.  Ultimately, if J1459$-$3852 or J1521$-$0607 are affiliated with Thamnos, this would suggest that the Thamnos progenitor experienced early $r$-process nucleosynthesis and was able to retain the ejecta.  This scenario is consistent with the estimated mass of $\lesssim$ 5$\times$10$^6 M_{\odot}$ from \citet{2019A&A...631L...9K}.
\end{description}

Ultimately, more high-resolution spectroscopic observations of metal-poor Thamnos stars are needed to determine whether these three stars could be associated with the stream.

\section{Summary}\label{sec:summary}
This paper presents an in-depth analysis of portrait spectra for three metal-poor, non-carbon-enhanced stars identified by the RPA, including cosmo-chronometric age dating with \ion{Th}{2} for two stars. %All three stars are found to be $\alpha$-enhanced.  
\cms{This portrait analysis presents abundance measurements or upper limits for 29 elements, significantly increasing the number of elements from previous RPA snapshot analyses of these stars.}  The detailed chemical abundances in this work confirm that 2MASS J1459$-$3852 and J1521$-$0607 are $r$-II stars, but 2MASS J1944$-$4039 is found to be not $r$-process-enhanced, with a [Eu/Fe] ratio just below the threshold for $r$-I status.  

\cms{This paper presents the first detections of Mo, Ru, Rh, Pd, and Ag in these stars, allowing examination of the amount of fission fragments in these stars.  The three stars show different relative amounts of these lighter $r$-process elements, consistent with the variations in [Eu/Fe].}  The differing $r$-process enrichment amongst the three stars may therefore be due to different contributions of fission fragments, as noted in other RPA papers, where the $r$-II stars show enhancement in certain elements (like Pd and Ag), relative to 2MASS J1944$-$4039.  

\cms{The paper also presents Th abundances for two stars and an upper limit on the third, as well as an upper limit on U for one star.  Detections of these radioactive elements with long half lives allow cosmo-chronometric age dating.}  The two $r$-II stars are also found to have \ion{Th}{2} abundances that indicate old ages for the $r$-process material.

All three stars have previously been \cms{found to have retrograde orbits, suggesting origins in an accreted dwarf galaxy.  Previous kinematic clustering analyses also found the stars to be in} CDTGs or DTGs, depending on the stellar samples used for clustering analyses.  The disparate chemical abundance ratios of the three stars \cms{found in this paper} suggest that the three stars likely did not form in the same environment, \cms{and are unlikely to be associated with a common CDTG or DTG}.  In particular, the relative lack of fission fragments in J1944$-$4039 shows that it received different $r$-process material from J1459$-$3852.  Individually, each star has a chemical signature that is consistent with origins in a low-mass dwarf galaxy.  Previous papers have suggested a link to the Thamnos substructure, though more work is needed to assess whether any of these three stars can indeed be linked to that stream.

%\clearpage
\acknowledgements 
The authors thank the anonymous reviewer for suggestions that improved the manuscript.
J.A. and C.M.S. acknowledge support from the National Science Foundation grant AST 2206379.
T.C.B. acknowledges partial support from grants PHY 14-30152; Physics Frontier Center/JINA Center for the Evolution of the Elements (JINA-CEE), and OISE-1927130; The International Research Network for Nuclear Astrophysics (IReNA), awarded by the US National Science Foundation, and DE-SC0023128; the Center for Nuclear Astrophysics Across Messengers (CeNAM), awarded by the U.S. Department of Energy, Office of Science, Office of Nuclear Physics. 
The work of V.M.P. is supported by NOIRLab, which is managed by the Association of Universities for Research in Astronomy (AURA) under a cooperative agreement with the U.S. National Science Foundation.
I.U.R.\ acknowledges support from the U.S.\ National Science Foundation (AST~2205847).
T.T.H\ acknowledges support from the Swedish Research Council (VR 2021-05556). 
R.E. acknowledges support from NASA Astrophysics Theory Program
grant 80NSSC24K0899.

\software{IRAF (\citealt{tody1986,tody1993,fitzpatrick2025}), MOOG (v2017; \citealt{1973ApJ...184..839S,2011AJ....141..175S}), linemake (\url{https://github.com/vmplacco/linemake})}

\footnotesize{}

\end{document}